\documentclass[
amssymb, %arxiv ON
amsmath, %
floatfix, %
twocolumn, %
reprint, %
superscriptaddress, %
prb, %
aps, %
nofootinbib,%
nobibnotes,%
longbibliography, %
lengthcheck, %
final, %
]{revtex4-1} %arxiv OFF
\renewcommand{\thispagestyle}[1]{}
\usepackage[T1]{fontenc}
\usepackage[utf8]{inputenc}
\usepackage{graphicx}
\usepackage{latexsym}
\usepackage{amsthm}
\usepackage[low-sup]{subdepth}
\usepackage{soul}
\usepackage{color}
\usepackage{mathtools}
\mathtoolsset{showonlyrefs}
\usepackage{notes2bib}

\usepackage[]{placeins}

\usepackage[dvipsnames]{xcolor}

\usepackage[]{newtxtext} %arxiv comment
\usepackage{bm}

\usepackage{floatrow}
\usepackage{siunitx}
\mathchardef\mhyphen="2D
\usepackage{hyperref}
\hypersetup{
	colorlinks,
	linkcolor={blue!90!black},
	citecolor={blue!90!black},
	urlcolor=	{blue!90!black}
}
\let\oldeqref\eqref
\renewcommand*{\eqref}[1]{%
	\hyperref[#1]{\oldeqref{#1}}%
}

\usepackage{ragged2e}

\newcommand{\tabref}[1]{Tab.~\ref{tab:#1}}
\newcommand{\figref}[1]{Fig.~\ref{fig:#1}}
\newcommand{\subfigref}[2]{Fig.~\hyperref[fig:#1]{\ref*{fig:#1}(#2)}}
\newcommand{\subfigsref}[3]{Figs.~\hyperref[fig:#1]{\ref*{fig:#1}(#2)}-\hyperref[fig:#1]{\ref*{fig:#1}(#3)}}

\definecolor{cbred}{HTML}{e31a1c}
\definecolor{cbgreen}{HTML}{33a02c}
\definecolor{cbblue}{HTML}{176aa7}
\definecolor{cborange}{HTML}{ff7f00}
\definecolor{cbviolet}{HTML}{6a3d9a}

\definecolor{tomimoto}{HTML}{66c2a5}
\definecolor{gong}{HTML}{e78ac3}

\newcommand*{\kp}{\bm{k}{\cdot}\bm{p}}
\newcommand*{\mfrac}[2]{#1/#2}

\DeclarePairedDelimiter\lr{\lparen}{\rparen}

\DeclarePairedDelimiterX{\comm}[2]{\lbrack}{\rbrack}{#1, #2}

\DeclarePairedDelimiterX{\braket}[2]{\langle}{\rangle}{#1\delimsize\vert #2}
\DeclarePairedDelimiterX{\ketbra}[2]{\rvert}{\lvert}{#1 \delimsize\rangle\!\delimsize\langle #2}
\DeclarePairedDelimiterX{\matrixel}[3]{\langle}{\rangle}{#1 \delimsize\vert #2 \delimsize\vert #3}

\newcommand{\mr}[1]{\mathrm{#1}}

\begin{document}

	\title{Optical and electronic properties of low-density InAs/InP quantum dot-like structures devoted to single-photon emitters at telecom wavelengths}
	
	\author{P.~Holewa}
\affiliation{%
	Laboratory for Optical Spectroscopy of Nanostructures, %
	Department of Experimental Physics, %
	Faculty of Fundamental Problems of Technology, %
	Wroc\l{}aw University of Science and Technology, %
	Wybrze\.ze Wyspia\'nskiego 27, %
	50-370 Wroc\l{}aw, %
	Poland%
}

\author{M.~Gawe{\l}czyk}
\email{michal.gawelczyk@pwr.edu.pl}
\affiliation{Department of Theoretical Physics, Faculty of Fundamental Problems of Technology, Wroc\l{}aw University of Science and Technology, 50-370 Wroc\l{}aw, Poland}
\affiliation{%
	Laboratory for Optical Spectroscopy of Nanostructures, %
	Department of Experimental Physics, %
	Faculty of Fundamental Problems of Technology, %
	Wroc\l{}aw University of Science and Technology, %
	Wybrze\.ze Wyspia\'nskiego 27, %
	50-370 Wroc\l{}aw, %
	Poland%
}

\author{C.~Ciostek}
\altaffiliation[Currently at ]{
	Experimental Physics 3, %
	University of W{\"u}rzburg, %
	Am Hubland, %
	97074 W{\"u}rzburg, %
	Germany%
}
\affiliation{%
	Laboratory for Optical Spectroscopy of Nanostructures, %
	Department of Experimental Physics, %
	Faculty of Fundamental Problems of Technology, %
	Wroc\l{}aw University of Science and Technology, %
	Wybrze\.ze Wyspia\'nskiego 27, %
	50-370 Wroc\l{}aw, %
	Poland%
}

\author{P.~Wyborski}
\affiliation{%
	Laboratory for Optical Spectroscopy of Nanostructures, %
	Department of Experimental Physics, %
	Faculty of Fundamental Problems of Technology, %
	Wroc\l{}aw University of Science and Technology, %
	Wybrze\.ze Wyspia\'nskiego 27, %
	50-370 Wroc\l{}aw, %
	Poland%
}

\author{S.~Kadkhodazadeh}
\affiliation{
	DTU Nanolab -- National Centre for Nano Fabrication and Characterization, Technical University of Denmark, Kongens Lyngby DK-2800, Denmark
}%

\author{E.~Semenova} \email{esem@fotonik.dtu.dk}
\affiliation{
	DTU Fotonik, Technical University of Denmark, Kongens Lyngby DK-2800, Denmark
}%

\author{M.~Syperek} \email{marcin.syperek@pwr.edu.pl}
\affiliation{%
	Laboratory for Optical Spectroscopy of Nanostructures, %
	Department of Experimental Physics, %
	Faculty of Fundamental Problems of Technology, %
	Wroc\l{}aw University of Science and Technology, %
	Wybrze\.ze Wyspia\'nskiego 27, %
	50-370 Wroc\l{}aw, %
	Poland%
}

	\begin{abstract}
		Due to their band-structure and optical properties, InAs/InP quantum dots (QDs) constitute a promising system for single-photon generation at third telecom window of silica fibers and for applications in quantum communication networks.
		However, obtaining the necessary low in-plane density of emitters remains a challenge.
		Such structures are also still less explored than their InAs/GaAs counterparts regarding optical properties of confined carriers.
		Here, we report on the growth via metal-organic vapor phase epitaxy and investigation of low-density InAs/InP QD-like structures, emitting in the range of $\SIrange{1.2}{1.7}{\micro\meter}$, which includes the S, C, and L bands of the third optical window.
		We observe multiple photoluminescence (PL) peaks originating from flat QDs with height of small integer numbers of material monolayers.
		Temperature-dependent PL reveals redistribution of carriers between families of QDs.
		Via time-resolved PL, we obtain radiative lifetimes nearly independent of emission energy in contrast to previous reports on InAs/InP QDs, which we attribute to strongly height-dependent electron-hole correlations.
		Additionally, we observe neutral and charged exciton emission from spatially isolated emitters.
		Using the 8-band $\kp$ model and configuration-interaction method, we successfully reproduce energies of emission lines, the dispersion of exciton lifetimes, carrier activation energies, as well as the biexciton binding energy, which allows for a detailed and comprehensive analysis of the underlying physics.
	\end{abstract}
	
	\maketitle
	
	\section{Introduction}
	Semiconductor epitaxial self-assembled quantum dots (QDs) have attracted attention for decades. Among many factors, this interest is driven by their ability to emit single photons of high purity and indistinguishability \cite{Aharonovich2016,Senellart2017, Scarpelli2019}, and by the possibilities of their integration with existing semiconductor platforms. The proposed real-world applications required further development of QDs to meet specific requirements, e.\,g. emission wavelength or spatial isolation of an emitter. A currently vital example is a non-classical single photon source desired for quantum-secured data transfer and quantum information processing in open systems exploiting the existing architecture of mid- and long-haul silica- fiber-based optical networks. To achieve this specific application target, a QD should emit in the low-loss third spectral window of silica fibers (\SI{1.46}--\SI{1.625}{\micro\meter}) commonly divided into three subbands: S ($\SIrange{1460}{1530}{\nano\metre}$), C ($\SIrange{1530}{1565}{\nano\metre}$), and L ($\SIrange{1565}{1625}{\nano\metre}$).
	This spectral window can be reached by employing an InGaAs metamorphic layer approach for the well-known InAs/GaAs QD structure \cite{Semenova2008}. Very recently, QDs made by this technique have shown high fidelity photon entanglement \cite{Olbrich2017ent}. However, this growth method is technologically demanding and still prone to structural defects \cite{Seravalli2011}, which obstructs its migration to the realm of applications.
	
	Emission in the third telecom band can naturally be achieved by using InAs QDs grown on the InP substrate without applying additional strain engineering to the band gap. However, due to more than two times smaller lattice-mismatch in comparison to standard InAs/GaAs QDs, the InAs/InP dots form dense arrays of flat and sizable in-plane material islands. Consequently, the in-plane spatial isolation and quality of such dots are hardly achievable but still indispensable for a single-photon source application.
	
	Up to now, the reduced planar density of InAs/InP QDs emitting in the targeted spectral range has been achieved in molecular beam epitaxy (MBE) through modifications of the growth scheme or post-growth annealing of the structure \cite{Kubota2010,Benyoucef2013,Yacob2014}. A similar approach has been used in metal-organic vapor phase epitaxy (MOVPE), where the combination of annealing with double-capping was applied to enhance the uniformity of QD sizes \cite{Takemoto2004a,Leavitt2015}. Both techniques have yielded good optical quality InAs/InP QD-based emitters \cite{Benyoucef2013, Musial2019, Miyazawa2016}. Here, we explore the MOVPE growth of InAs/InP QD-like nanostructures, modified by introducing the additional interruption step leading to formation of sparse islands emitting light within the third telecom band. The strain-driven formation of islands (the Stranski-Krastanov mechanism) from the InAs(P) layer is partially suppressed by high V/III ratio to ensure low surface density of QDs even slightly below \SI{e9}{\centi\metre\tothe{-2}}.
	This results in good spatial isolation of single emitters allowing for examination of their optical properties.
	
	Apart from their relevance for applications, InAs QDs grown on InP itself or on InP lattice-matched alloys, e.\,g. InGaAlAs with various cation proportions, are interesting for fundamental research, as they provide carrier confinement distinct from the one known from well-explored and described GaAs-based systems.
	Recent studies have provided explanation of some of their properties, such as partially polarized emission \cite{Musial2012}, exciton recombination characterized by two distinct lifetimes in the case of asymmetric dots \cite{Gawelczyk2017} or the structure of excited states \cite{Gawelczyk2019}. They have also raised a question regarding the exciton confinement regime \cite{Gawelczyk2018, Dusanowski2017a}.
	These results were, however, obtained mostly for very dense ensembles of elongated QDs \cite{Sauerwald2005}, and thus characterization of carrier confinement and the resulting optical properties of low-density QDs, growing in a different geometry such as those considered here, is needed. In this work, we combine both goals and provide a comprehensive description of this specific low-density QD-like system, focusing both on its relevance for potential applications and on fundamental properties of carrier confinement leading to the observed properties of emission.
	%\esst{We establish that} 
	
	The investigated InAs/InP QDs are formed on a 3 ML thick InAs(P) wetting layer (WL) and have truncated pyramidal shape with hexagonal base. The ensemble photoluminescence (PL) from the dots is concentrated in multiple narrow bands corresponding to QDs with height of consecutive integer numbers of MLs of material, according to our calculation. By fine-tuning the agreement of calculated exciton energies with positions of PL peaks, we establish that QDs in-plane size scales linearly with their height.
	Additionally, from time-resolved PL studies, we assess exciton lifetimes on the order of \SI{1}{\nano\second} with a very weak dispersion of values.
	While in general a dispersive character of the latter is expected due to strong scaling of radiative recombination rates with emitted wavelength \cite{Karrai2003}, we find theoretically that the virtually energy-independent values measured here result from the interplay between the said general trend and a strong change in the strength of Coulomb electron-hole correlations, known to enhance exciton-light coupling, with increasing QD height.
	We observe also a thermally-driven redistribution of carriers, exhibited in thermal quenching of PL from the highest-energy peaks and combination of enhancement and quenching for the others.
	For the latter the extracted activation energies correspond well to those calculated by us for escape of electrons to the WL, while for the other we rather deal with excitation of whole excitons.
	At the same time, the energies corresponding to PL enhancement for each of the bands correspond to energies of thermal escape of single carriers from QDs forming the higher-energy neighboring band, which illustrates the migration of carriers between families of QDs.
	Finally, by means of high-resolution microphotoluminescence (\si{\micro}PL) measurements of the samples with fabricated sub-micrometer mesas, we observe emission lines originating from recombination of neutral and charged excitonic complexes within single QDs.
	Charged excitons (CX) are observed in each of the subbands of the third telecom window with emission persisting up to $T = \SI{70}{\kelvin}$.
	
	The paper is organized as follows.
	In Sec.~\ref{sec:system}, we describe the investigated structure and experimental methods.
	Next, in Sec.~\ref{sec:theory} we describe the formalism employed for theoretical calculations.
	In Sec.~\ref{sec:results}, we report experimental results and provide explanations based on theoretical calculations.
	Finally, we summarize our findings in Sec.~\ref{sec:summary}.
	
	\section{Investigated structure and optical experiments}\label{sec:system}
	In this Section, we discuss the QDs growth method, and the morphological characteristics of the structures under consideration, and describe the experimental methods used.
	
	\subsection{Investigated structure}
	\begin{figure}[!tb] %
		\begin{center} %
			\includegraphics[width=\columnwidth]{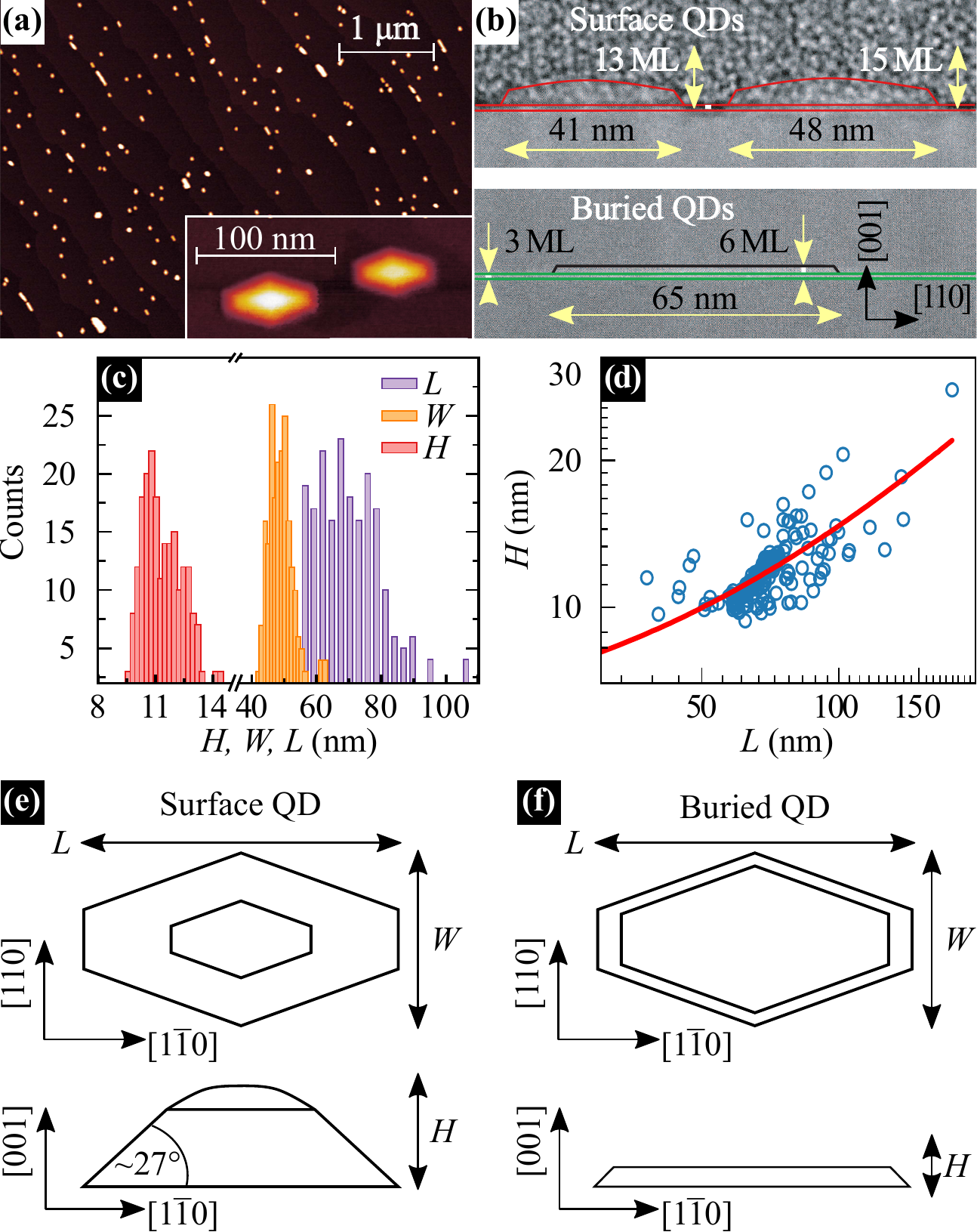} %
		\end{center} %
		\vspace{-0.8em}\caption{\label{fig:Structural}%
			(a) AFM picture of $5 \times 3.5$ \si{\micro\meter\squared} sample area with surface QDs.
			Inset: Magnified part revealing geometry of single QDs.
			(b) TEM image of the sample cross-section with surface and buried QDs.
			(c) Histograms of width $W$, length $L$, and height $H$ of surface QDs based on the AFM data.
			(d) Dependence of surface QDs height on their in-plane size based on AFM data (symbols) with a linear trend (line).
			(e) and (f) Assumed shapes for the surface and buried QDs, respectively.}
	\end{figure}
	The structure was grown in the low-pressure MOVPE TurboDisc\textregistered\ reactor using standard precursors: arsine (AsH$_3$), phosphine (PH$_3$), tertiarybutylphosphine (TBP) and trimethylindium (TMI).
	The growth sequence started from the deposition of a \SI{0.5}{\micro\meter} thick InP buffer layer on an (001)-oriented InP substrate at \SI{600}{\degreeCelsius}.
	Subsequently, the temperature was decreased down to \SI{480}{\degreeCelsius} and stabilized first under TBP and afterwards AsH$_3$ ambient. 
	A layer of QDs on a wetting layer (WL) was formed after deposition of a 1.04 ML thick InAs layer at TMIn and AsH$_3$ flow rates of \SI{12.2}{\micro \mole\per\minute} and \SI{4461}{\micro\mole\per\minute}, respectively, with the V/III ratio of 366.
	After that, a \SI{6}{\second} long growth interruption was introduced prior to the deposition of a \SI{10}{\nano\meter} thick InP cap layer.The temperature was then increased to \SI{600}{\degreeCelsius} and \SI{30}{\nano\meter} of InP was deposited. Finally, an array of surface QDs was deposited using the same growth parameters to form surface QDs for characterization by atomic force microscopy (AFM).

 The structural and morphological characteristics of the structures are summarized in \figref{Structural} and further elaborated in the Supplemental Information \cite{SI}.\vphantom{\cite{Hytch1998}} \subfigref{Structural}{a} shows a \SI{5 x 3.5}{\micro\meter} areal scan of surface QDs measured with the AFM.
	The dots are relatively large and slightly elongated in $[1\overline{1}0]$ direction and are sparsely scattered with a surface density of $\mathbin{\lesssim}\SI{e9}{\centi\metre\tothe{-2}}$.
	Statistical analysis of their geometry presented in \subfigref{Structural}{c} showed length $L$ in the range of $\SIrange{55}{85}{\nano\meter}$, while width $W$ is spread between \SI{45}{\nano\meter} and \SI{55}{\nano\meter}.
	This gives an average in-plane aspect ratio of $\sim 1.25$.
	The median QD height $H$ is relatively large and exceeds \SI{9}{\nano\metre}.
	Setting the aspect ratio of $L/W = 1.25$, we calculate $L$ from measured QD areas and check their correlation with $H$ by plotting in \subfigref{Structural}{d} the dependence $H(L)$ in the form of a scatter plot.
	Based on this we deduce a linear dependence between $H$ and $L$ for surface QDs, which is plotted to guide the eye with a solid line.
	
	The magnified image of surface QDs shown in the inset of \subfigref{Structural}{a} reveals their diamond-like shape ended from top by a dome, which we schematically present in \subfigref{Structural}{e}.
	Although these dots are optically inactive, we assume that their in-plane dimensions are similar to their buried counterparts, being the source of strong optical response.
	
	A high-angle annular dark-field scanning
	transmission electron microscopy (HAADF STEM) of the structure is presented in \subfigref{Structural}{b}, providing a cross-sectional view in $[1\overline{1}0]$ crystallographic direction. The contrast in HAADF STEM images is primarily related to the atomic number and can thus be used directly to measure the dimensions of the QDs \cite{Kadkhodazadeh2013}.
	The upper panel reveals surface QDs, while the bottom one shows a single, buried dot.
	Although some similarities between both types of dots are discernible, significant differences between them can also be observed.
	Both types of dots are of similar size in the growth plane.
	However, the height of buried QDs is much smaller than of the surface ones.
	Additionally, the buried ones are flat on top [\subfigref{Structural}{f}], in contrast to dome-shaped surface dots [\subfigref{Structural}{e}].
    These differences most probably result from longer growth interruption after QDs nucleation for surface QDs. In this case, the material may be partially redistributed from the WL to the top of QDs, driven by strain relaxation. However, we can assume that the lateral sizes of surface dots are the same as of buried dots.
	The buried QD shown protrudes by 6\:ML from the WL top surface. 
	We estimate that QDs are purely InAs from STEM scans and extracted fractional lattice spacing changes which can be found in the Supplemental Information \cite{SI}.
	
	\subsection{Optical experiments}
	For time-integrated PL, photoluminescence excitation (PLE), and time-resolved PL (TRPL) experiments the structure was held in a helium closed-cycle refrigerator allowing for control of temperature in the range of $\SIrange{10}{300}{\kelvin}$.
	The sample was excited by a train of \SI{\sim2}{\pico\second} long pulses generated by the optical parametric oscillator (OPO) providing a photon energy in the range of $\SIrange{0.9}{1.2}{\electronvolt}$.
	The pulse generation frequency was either \SI{76}{\mega\hertz} or lower.
	The light emitted from the sample was collected in a standard lens-based far-field optical setup and dispersed by a 0.3-m-focal-length monochromator.
	Time-integrated PL was measured in a wide spectral range of $\SIrange{0.5}{1.1}{\electronvolt}$ via the lock-in technique at the reference modulation frequency of \SI{2}{\kilo\hertz}, using a thermoelectrically cooled InGaAs-based single-channel detector.
	The same experimental setup was employed for the PLE measurements.
	In this case, the monochromator was fixed to a certain detection energy ($E_{\mr{det}}$) while the structure was excited by scanning the photon energy across the OPO's spectral range.
	The TRPL was measured by a time-correlated single photon counting method.
	Photons were spectrally filtered by the monochromator and subsequently directed onto the NbN superconducting detector.
	The multichannel event timer was synchronized to the pulse train to produce photon event statistics.
	The overall temporal resolution of the TRPL setup was \SI{\sim80}{\pico\second}.
	
	For the PL with high spatial resolution (\si{\micro}PL) the sample was kept in a He-flow cryostat allowing for control of the sample temperature in the range of $\SIrange{5}{300}{\kelvin}$. Sample was excited by the continous wave (CW) semiconductor laser, and the emission was collected by a near-infrared-optimized microscope objective with high numerical aperture of $\mathrm{NA}=0.4$.
	The \si{\micro}PL signal was spectrally analyzed by a 1-m-focal-length monochromator and registered by a liquid-nitrogen-cooled InGaAs-based linear detector operating in the spectral range from \SI{1.24}{\electronvolt} down to \SI{0.75}{\electronvolt}.
	Polarization properties of emitted light were analyzed by rotating the half-wave retarder mounted before a fixed high-contrast-ratio ($10^6{:}1$) linear polarizer placed in front of the monochromator's entrance.
	
	\section{Theory}\label{sec:theory}
	In this Section, we describe the theoretical framework used to model investigated QDs and their electronic and optical properties.
	
	\subsection{Modeling of QDs}
	The modeling of nanostructures in question was preceded by evaluation of our initial assumption on their height varying by single MLs and by establishing the thickness of the WL.
	For this, numerical calculations within the eight-band $\kp$ method with use of the commercially available \textit{nextnano} software have been performed \cite{Birner2007}.
	As a result, via comparison with experimental data, a WL thickness of 3\:MLs has been determined.
	
	To model the buried QDs we have used the available structural data presented in Sec.~\ref{sec:system}.
	While the in-plane shape is not well-established, we followed the premise derived from surface dots and assumed an elongated hexagonal base with aspect ratio of $L/W = 5/4$.
	The height, based on the cross-sectional data, is of a few MLs of InAs.
	Additionally, from the spacing of PL peaks we form an assumption that the dots are of height equal to consecutive integer numbers of MLs.
	Thus, we have modeled a series of such QDs protruding by $N$~MLs above the WL of 3~ML thickness, for $N=\numrange{1}{9}$.
	For each height, we simulated QDs with in-plane sizes varied uniformly from $\SI{15}{\nano\meter}\times\SI{12}{\nano\meter}$ up to $\SI{60}{\nano\meter}\times\SI{48}{\nano\meter}$ in 7 linear steps.
	Again following the properties of surface QDs, we set the inclination of QD side walls to $\SI{27}{\degree}$, which is however of little relevance in view of their small height.
	Regarding the material composition, we assumed that QDs and WL are of pure InAs.
	
	\subsection{Single-particle states and complexes}
	\begin{table}[tb!]
		\newcommand*{\cv}{\cite{Vurgaftman2001}}
		\newcommand*{\cw}{\cite{Winkler2003}}
		\newcommand*{\cs}{\cite{Saidi2010}}
		\newcommand*{\ct}{\cite{Tse2013}}
		\newcommand*{\cc}{\cite{CaroPRB2015}}
		\newcommand*{\ca}{\cite{Amirtharaj1994}}
		\newcommand*{\cmsq}{${\mfrac{\mathrm{C}}{\mathrm{m}^2}}$}
		\newcommand*{\intEg}{$^{-0.13}_{+1.31x}$}
		\begin{tabular*}{\columnwidth}{ @{\extracolsep{\fill}} ccccc}
			\hline\hline\rule{0pt}{1.1em} 
			Parameter & Unit & InAs & InP & Source\\\hline
			$a_\mathrm{}$ & \AA &6.06 &5.87 & \cv \\
			$E_\mathrm{g}$ & eV &0.417 &1.42 & \cv \\
			VBO & eV &-0.59 &-0.94 & \cv \\
			$E_\mathrm{p}$ & eV &21.5 &20.7 & \cv \\
			$m_\mathrm{e}^{*}$ & &0.0229 &0.0803 & \cw \\
			$\Delta_\mathrm{}$ & eV &0.39 &0.11 & \cv \\
			$\gamma_\mathrm{1}$ & &20.4 &4.95 & \cw \\
			$\gamma_\mathrm{2}$ & &8.3 &1.65 & \cw \\
			$\gamma_\mathrm{3}$ & &9.1 &2.35 & \cw \\
			$e_\mathrm{14}$ & \cmsq &-0.111 &0.016 & \cc \\
			$B_\mathrm{114}$ & \cmsq &-1.17 &-1.54 & \cc \\
			$B_\mathrm{124}$ &\cmsq &-4.31 &-3.62 & \cc \\
			$B_\mathrm{156}$ & \cmsq &-0.46 &-1.02 & \cc \\
			$C_\mathrm{k}$& eV\,\AA &-0.0112 &-0.0144 & \cw \\
			$a_\mathrm{c}$ & eV &-5.08 &-6.0 & \cv \\
			$a_\mathrm{v}$ & eV &1.0 &0.6 & \cv \\
			$b_\mathrm{v}$ & eV &-1.8 &-2 & \cv \\
			$d_\mathrm{v}$ & eV &-3.6 &-5 & \cv \\
			$c_\mathrm{11}$ & GPa &833 &1011 & \cv \\
			$c_\mathrm{12}$ & GPa &453 &561 & \cv \\
			$c_\mathrm{44}$ & GPa &396 &456 & \cv \\
			$\varepsilon_\mathrm{r}$ & &14.6 &12.4 & \cw\\
			\hline\hline
		\end{tabular*}\vspace{-0.7em}
		\vspace{-0.4em}\caption{\label{tab:material} Material parameters used in the modeling of nanostructures and calculation of single-particle and exciton states.}
	\end{table}
	The lattice mismatch between InAs and InP is only approximately \SI{3}{\percent}, nonetheless we have calculated the structural strain within the standard continuous elasticity theory via minimization of elastic energy.
	Next, piezoelectric field resulting from the presence of shear strain in a noncentrosymmetric material has been calculated up to second order terms in strain-tensor elements.
	With this, to calculate the electron and hole eigenstates and then states of their complexes, we have used a numerical implementation \cite{GawareckiPRB2014} of the eight-band $\kp$ theory with envelope-function approximation \cite{Burt1992,Foreman1993}.
	To fix the spin configuration, a weak (\SI{20}{\milli\tesla}) magnetic field was used.
	The Dresselhaus spin-orbit interaction enters via perturbative terms in Hamiltonian blocks that couple the conduction band to valence bands, while the impact of structural strain is included via the standard Bir-Pikus Hamiltonian \cite{Bir1974}.
	The explicit form of the Hamiltonian and description of the numerical implementation may be found in Ref.~[\onlinecite{Gawarecki2018}], while the material parameters used are given in \tabref{material} along with their sources.
	Numerical diagonalization yields the conduction- and valence-band electron eigenstates in the form of pseudospinors with components holding the envelope functions for each of the subbands.
	Hole states were obtained by application of time reversal operation to valence-band electron states.
	
	Next, the calculated electron and hole single-particle states have been used to calculate the neutral- and charged-exciton, as well as biexciton states in the configuration-interaction approach.
	To this end, electron-hole Coulomb and phenomenological anisotropic exchange interactions have been diagonalized in the basis of $40\times 40$ electron-hole configurations, which yielded carrier-complex states expanded in the configuration-space basis.
	Finally, for each of the states the coupling to light has been evaluated in the dipole approximation, and the resultant dipole moments were used to calculate radiative lifetimes of carrier-complex states \cite{ThraenhardtPRB2002}.
	
	\section{Results and discussion}\label{sec:results}
	In this Section, we present both experimental and theoretical results.
	Their confrontation and mutual feedback are used as a base to discuss in detail the revealed properties of the system.
	
	\subsection{Emission from ensemble of QDs}
	\begin{figure}[!tb] %
		\begin{center} %
			\includegraphics[width=\columnwidth]{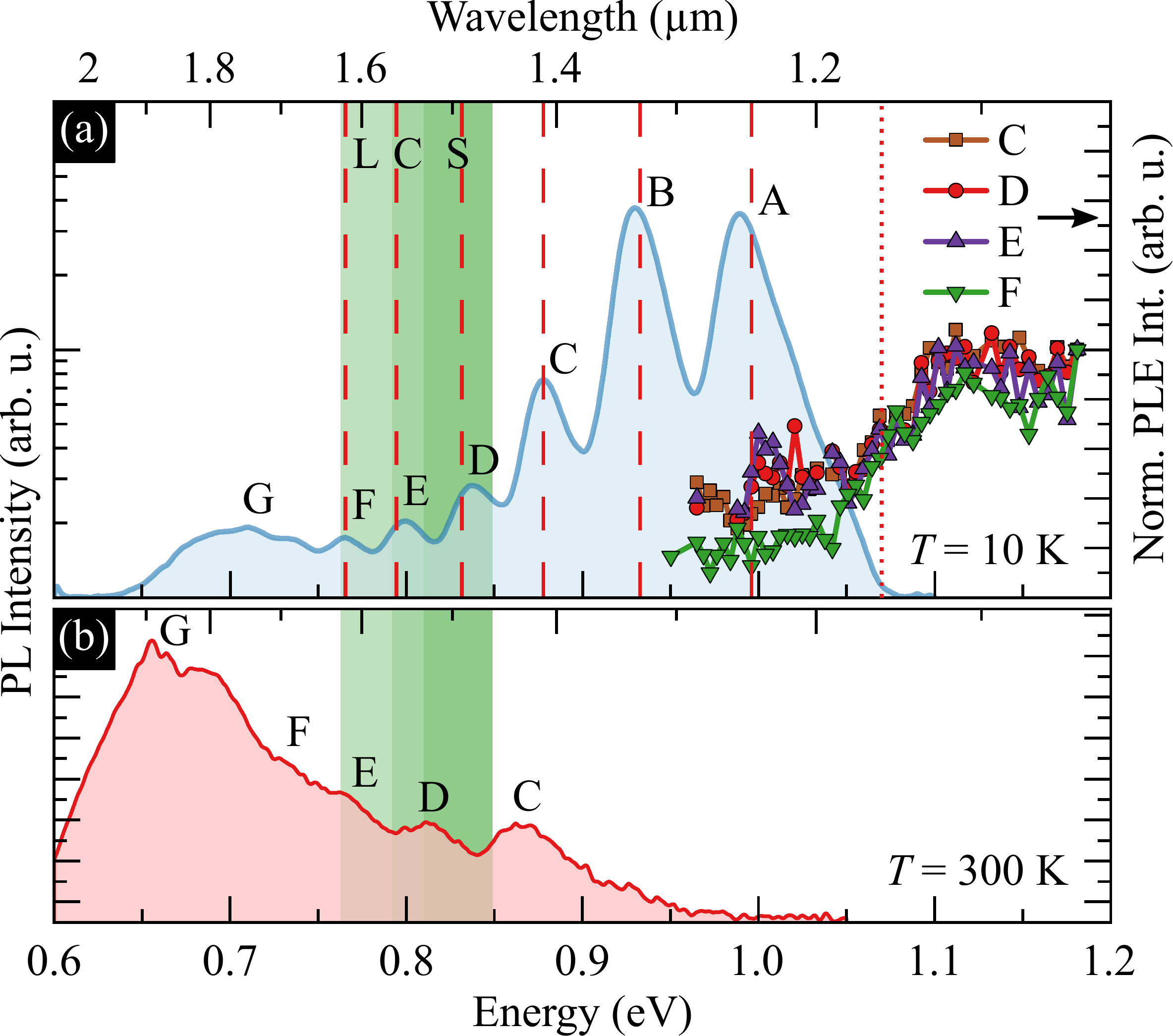} %
		\end{center} %
		\vspace{-0.8em}\caption{\label{fig:PL_PLE_spectrum}%
			(a) Low temperature ($T=\SI{10}{\kelvin}$) PL spectrum (solid line, log scale) and PLE signals for the ensemble of QDs obtained for collection from lines C-F (symbols).
			Vertical dashed lines mark theoretically calculated exciton energies for modeled QDs and dotted line marks the wetting layer absorption edge.
			Shaded areas represent the S- ($\SIrange{1460}{1530}{\nano\metre}$), C- ($\SIrange{1530}{1565}{\nano\metre}$) and L- ($\SIrange{1565}{1625}{\nano\metre}$) bands of the third telecom window.
			(b) PL spectrum measured at $T = \SI{300}{\kelvin}$.}
	\end{figure}
	We begin with presenting the results of PL measurements performed on the ensemble of QDs in question.
	These are presented in \subfigref{PL_PLE_spectrum}{a} where at $T=\SI{10}{\kelvin}$ a multimodal distribution of intensity with well-pronounced seven PL maxima (labeled A to G) is visible.
	The peaks spread across the energy range of $\SIrange{\sim0.65}{1.05}{\electronvolt}$, and, notably, those labeled with D, E and F coincide with the S, C and L bands of the third telecom window, respectively.
	Combining this observation with knowledge on the growth process, we conjecture that each of the PL bands corresponds to emission from a distinct family of QDs. Such multi-peaked PL was previously observed for InAs/InP QDs grown via chemical beam epitaxy (CBE) \cite{Gustafsson1995,Poole2001}, MBE \cite{Folliot1998,Berhane2001}, and MOVPE \cite{Sakuma2005,Lanacer2007}, as well as for InAs/GaAs QDs \cite{Guffarth2001}. Based on our initial theoretical estimations, to form such a pattern of bands in emission, QDs have to be of height varying by a single InAs ML ($\approx \SI{0.3}{\nano\metre}$), and each peak corresponds to a given QD height.
	Increase of temperature up to $T=\SI{300}{\kelvin}$ results in the PL pattern presented in \subfigref{PL_PLE_spectrum}{b}, which still reflects the multimodal QD-height distribution.
	Thus, the confinement for carriers is strong enough, so that they are still localized, as it could be expected for QDs of size and composition assumed here.
	
	Next, we focus on absorption properties of the dots, which are reflected in the PLE spectrum, shown with points in \subfigref{PL_PLE_spectrum}{a}.
	It has been measured for four families of QDs by fixing $E_{det}$ at their PL peak energy and registering the emission intensity while scanning the excitation energy.
	A strong absorption edge at about \SI{1.07}{\electronvolt} is present for all traces, which corresponds to the fundamental band gap of the WL.
	Moreover, since all traces are very similar, the WL beneath different QDs families apparently has a similar thickness and chemical composition.
	Theoretical calculations of the InAs/InP WL fundamental gap as a function of the WL thickness ($d_{WL}$) and its further comparison to the measured absorption edge leads to $d_{WL}\approx$ 3 ML.
	This value agrees with the WL thickness obtained from the cross-sectional TEM presented in \subfigref{Structural}{b}.
	
	The quantitative agreement between positions of PL peaks and calculated exciton energies, visible in \subfigref{PL_PLE_spectrum}{a}, has been achieved in the following way.
	As described in Sec.~\ref{sec:theory}, we have modeled a series of QDs with heights of consecutive integer numbers of InAs MLs and with varying in-plane size.
	This resulted in a smooth dependence of exciton ground-state energy on both varied QD dimensions.
	The latter could thus be interpolated via fitting of expected analytical dependence
	\begin{equation}\label{eq:energy-fit}
		E\lr*{H,L}
		\simeq \sum_{ij} a_{ij} \, H^{-i}\, L^{-j},
	\end{equation}
	where $a_{ij}$ are the fit coefficients, which nonnegative integer indices $i$ and $j$ do not exceed $i+j=2$.
	Next, we look for a smooth $L=f\lr{H}$ dependence.
	From the surface-QD structural data we draw an assumption that it may be linear [see \subfigref{Structural}{d}], thus we plug $L=A\,H+B$ into Eq.~\eqref{eq:energy-fit} and fit the resultant one-variable function of $H$ to the dependence of PL peaks positions on the assumed $\lr*{N+3}$\:ML height of corresponding QDs (a QD protrudes by $N$ MLs over the top of 3\:ML high WL).
	As a result, we obtain $L[\mr{nm}] = \lr*{3.2\pm0.5}\, N + \lr{ 28 \pm2}$, for which the final calculated energies are shown in \subfigref{PL_PLE_spectrum}{a}, in a more than satisfactory agreement with positions of all peaks.
	The final series of sizes for modeled QDs established here is then used for calculation of other quantities presented in the following subsections.
	
	\subsection{Temperature-driven effects}
	\begin{figure*}[!tb] %
		\begin{center} %
			\includegraphics[width=\columnwidth]{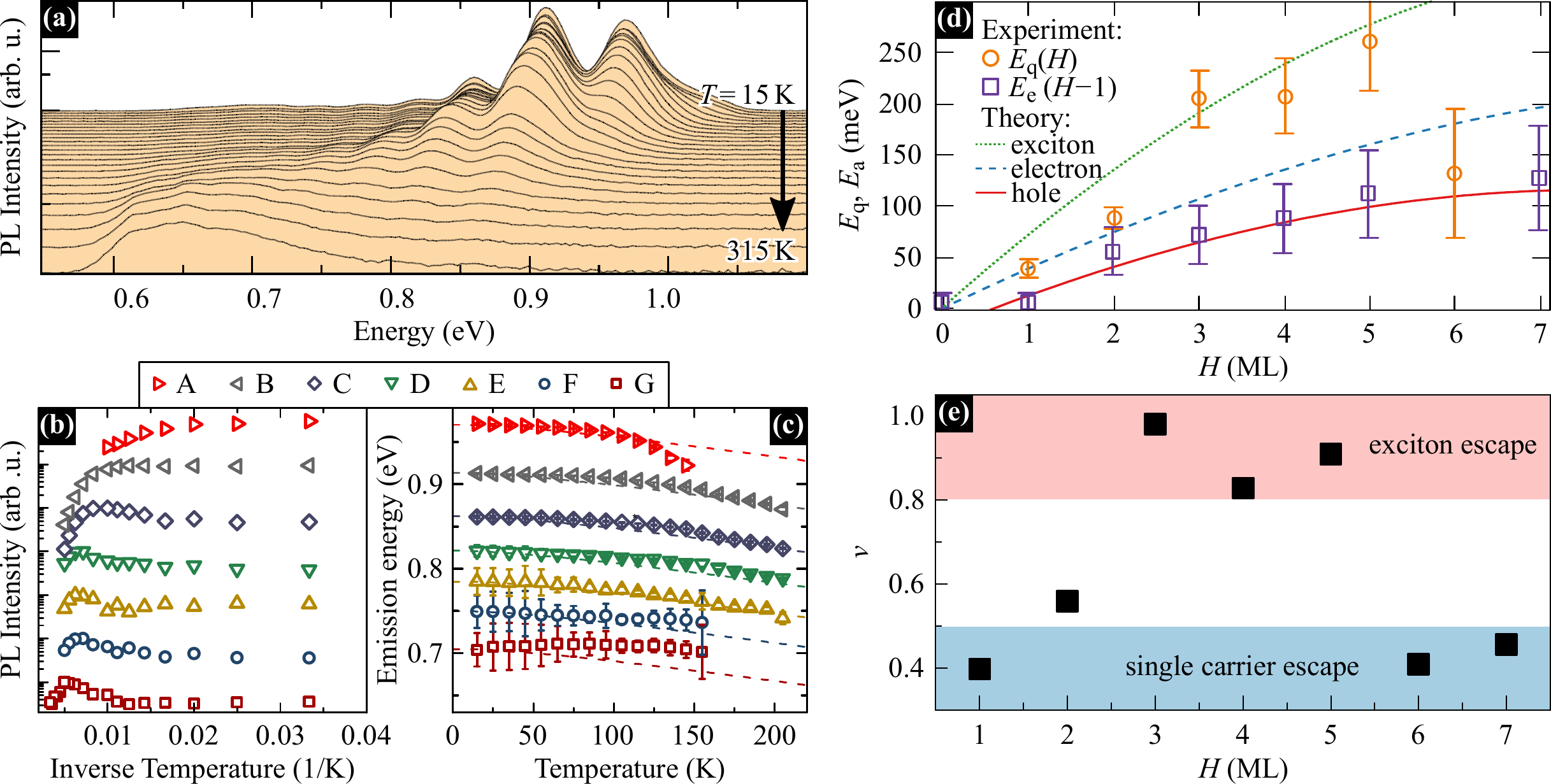} %
		\end{center} %
		\vspace{-0.8em}\caption{\label{fig:TempSeries_DTU}%
			Temperature dependence of PL emission from QDs.
			(a) Stacked PL spectra plotted in the Arrhenius form.
			(b) Temperature dependence of PL intensity for maxima A-G.
			(c) Thermal shift of emission energies (points) with Varshni lines: Eq.~\eqref{eq:varshni} (dashed lines).
			(d) Fitted activation energies for PL quench $E_{\mr{q}}$ (circles) and PL enhancement $E_{\mr{e}}$ (squares; shifted horizontally by one ML), and calculated energies for excitation to the WL for excitons (dotted line), electrons (dashed), and holes (solid) versus QD height.
			(e) Analysis of phenomenological parameter $\nu$.}
	\end{figure*}
	Here we move to the temperature-driven processes in the system.
	In \subfigref{TempSeries_DTU}{a}, we present the measured temperature-dependent PL spectra, where a strong redshift of emission is present, initially marked in \subfigref{PL_PLE_spectrum}{b}.
	Moreover, an overall intensity drop of emitted light is observed.
	We attribute this PL quenching to redistribution of carriers between QDs, during which part of them escapes from the low-dimensional traps completely.
	Such processes may result from differences in confinement potential depth for distinct QD families.
	They should involve thermal activation of carriers from QDs to the WL, their migration, and then re-trapping with the highest probability by larger QDs with deeper confining potential, where the carriers finally recombine.
	Such interpretation is supported by temperature dependence of intensities of each of the peaks plotted in the Arrhenius form in \subfigref{TempSeries_DTU}{b}.
	Here, the initial enhancement of PL intensity from the lower-energy PL bands (C--F) is observed up to intermediate temperature range.
	These QDs families correspond to deeper confinement of carriers.
	On the other hand, the highest-energy peaks exhibit only (band A) or mostly (B) a single process of PL quenching.
	This combination of observations illustrates how carriers are, on average, moved from lower to higher QDs.
	We note that the significant enhancement of PL intensities for QD families emitting in the third telecom window (families D--F) at above-cryogenic temperatures ($\mathbin{>}\SI{77}{\kelvin}$) is particularly advantageous from the point of view of possible applications as sources of single photons operating at elevated temperatures.
	
	In \subfigref{TempSeries_DTU}{c}, we show how the PL intensity enhancement and quenching are accompanied by partly suppressed thermal redshift of the low-energy peaks, and by enhanced one for the highest-energy band A.
	Nominally, for emission from the bulk material, a thermal redshift following the trend approximated by the Varshni relation is expected \cite{Varshni1967},
	\begin{equation}\label{eq:varshni}
		E_{\mr{g}}\lr*{T}
		= E_{\mr{g}}\lr*{0}-\frac{\alpha \, T^2}{T+\beta},
	\end{equation}
	where $E_{\mr{g}}$ is the band gap energy and parameters for InAs are $\alpha=\SI{2.76e-4}{\electronvolt\per\kelvin\tothe{2}}$, $\beta=\SI{93}{\kelvin}$.
	Corresponding curves are plotted with dashed lines.
	However, the Varshni formula takes into account only the thermal change of the semiconductor  band gap.
	The minor blueshift from these predictions observed here is most probably related to the state filling effect: at higher temperature carriers are redistributed within each of the QD-family ensembles so that higher energy-emitting QDs get occupied, i.\,e., those smaller in-plane and thus characterized by weaker confinement.
	Therefore, the high-energy tails of QDs distributions become comparatively more optically active as the temperature increases.
	
	For QDs with the shallowest confining potential (band A), emission energy initially follows the Varshni relation, but around \SI{120}{\kelvin} it considerably deviates due to fast depletion of high-energy QDs associated with strong PL quench.
	For families B--E, the emission energy follows formula \eqref{eq:varshni} within the entire temperature range.
	The different behavior for maximum G exhibiting almost no energy shift may indicate that structures forming this peak do not form QD family analogous to QDs A--F, whose well-defined central emission energies shift spectrally.
	This interpretation is supported by very broad emission from this maximum (see \figref{PL_PLE_spectrum}).
	Instead, the maximum G may be attributed to material islands where the size, and hence emission energy, change quasi-continuously.
	Additionally, in these islands we may expect a higher density of structural defects than in QDs families.
	
	To verify these initial qualitative speculations, PL intensity data for peaks A and B was fitted with a formula taking into account one activation process of energy $E_{\mr{q}}$ and relative rate $B_{\mr{q}}$ \cite{LambkinAPL1990}:
	\begin{equation}\label{ArrhEq}
		I\lr*{T} = \frac{I_0}{1+B_{\mr{q}} \exp\lr*{ -\frac{E_{\mr{q}}}{k_{\mr{B}}T} } },
	\end{equation}
	where $I_0$ is the PL intensity for $T\to 0$.
	For dots from families C--G, this analysis has to be extended to account for temperature-driven supply of carriers.
	Their recapturing and resulting PL enhancement is described by a characteristic energy, $E_{\mr{e}}$, corresponding to the process of carriers release from their reservoir and act as another source of occupation (apart from optical excitation) for the given bright state.
	Thus, we modify Eq.~\eqref{ArrhEq} in a straightforward manner by adding a state to the set of source kinetic equations from Ref.~\cite{LambkinAPL1990}, on the same footage as the given emitting state has been introduced, but with different activation energy to the common higher state (here: WL).
	With this a lengthy but simple to use formula is obtained, which accounts for exchange of carriers through the WL.
	Note that arbitrary changes made via educated guess in Eq.~\eqref{ArrhEq} that may be found in literature may lead to spurious values of extracted activation energies, if the form of modification does not comply with the nature of reservoir and its coupling with given state.
	Activation energies for PL quench and enhancement extracted in the described way are presented in \subfigref{TempSeries_DTU}{d} (points) versus the height given in the number of monolayers assigned to each of the peaks.
	Additionally, we plot the theoretically calculated energies needed for excitation of the electron, the hole, and the whole exciton to the WL ground state, taking the Coulomb coupling energies into account (lines).
	The results are obtained for the set of modeled QDs described above that led to agreement of exciton energies with maxima of PL bands.
	Comparing values of $E_{\mr{q}}$ to theoretical lines, we notice that also in this aspect peaks A and B behave differently than the few next ones.
	For these two, the activation energy related to quenching fits well to the value needed for escape of the electron from a QD.
	Then, for peaks C--E we rather deal with escape of whole excitons.
	Another change of behavior is present for the lowest-energy peaks, where a value between electron and hole escape is found, which possibly means that both these processes take place with similar probability.
	To understand the source of carriers that feed the thermally enhanced emission in the moderate temperature regime, we plot the values of $E_{\mr{e}}$ extracted via fitting, but we shift them by one ML with respect to the value for which they were observed.
	This helps us to notice that possibly the most effective redistribution takes place between QDs with 1~ML difference in height, i.\,e., that N-ML-high QDs are fed mostly by carriers that escape from QDs that are (N-1)-ML-high.
	Additionally, according to theoretical curves, this process is dominated by holes, which means that, while both carriers escape QDs, it is more probable for holes to be recaptured to another QD.
	This is intuitively in line with their higher effective mass and lower mobility.
	Thus, the assumption about migration of carriers between QDs from different families through the WL proves to be reasonable.
	
	The dominant PL quenching mechanisms assessed via fitting are also in agreement with those determined by calculation of the phenomenological parameter
	\begin{equation}
		\nu_i = \frac{E_{\mr{q}}^{(i)} }{ \Delta E_i},
	\end{equation}
	where $i$ labels QD families, and $\Delta E_i$ stands for the difference between transition energies in the QDs and in the WL, $E_{\mr{WL}}$.
	From PLE experiments we take $E_{\mr{WL}}=\SI{1.07}{\electronvolt}$ and achieve values presented in \subfigref{TempSeries_DTU}{e}.
	Generally, $\nu<0.5$ indicates a dominant role of a single-carrier escape and in this case $E_q$ corresponds to the lower of single-particle confinement energies (of either electron or hole) \cite{Gelinas2010}.
	For $\nu=0.5$ we expect correlated escape of electron-hole pairs \cite{Yang1997} and, finally, for $\nu\approx 1$ escape of excitons takes place \cite{Khatsevich2005}.
	
	\subsection{Photoluminescence dynamics}
	\begin{figure}[!tb] %
		\begin{center} %
			\includegraphics[width=\columnwidth]{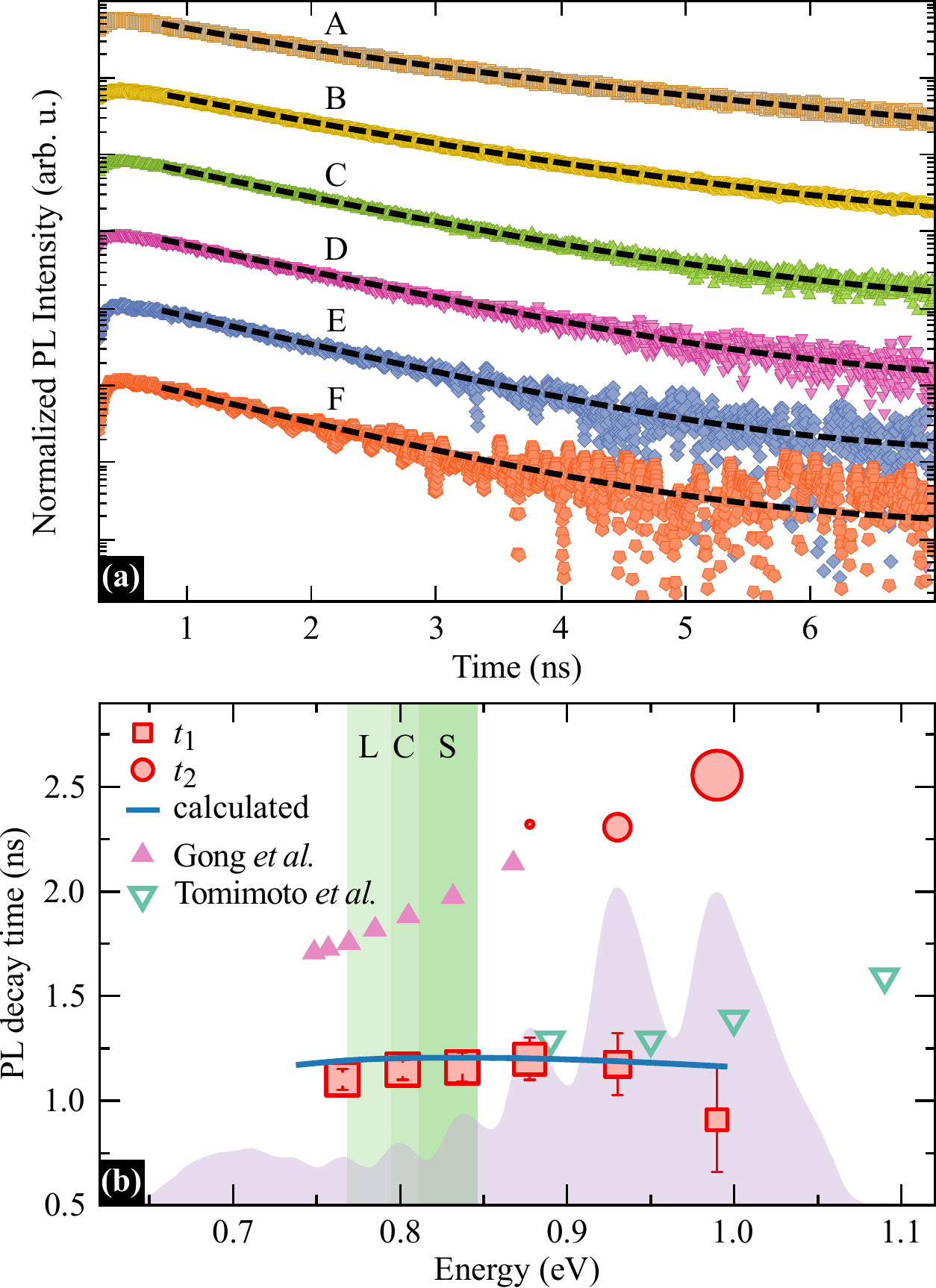}
		\end{center} %
		\vspace{-0.8em}\caption{\label{fig:Decays_and_MEM}%
			(a) Time-resolved photoluminescence traces for all PL maxima (points) along with reconstructed curves (lines).
			(b) Experimentally obtained PL decay times (open symbols; values correspond to means of fitted lifetime distributions; point size marks the component amplitude) compared with calculated radiative exciton lifetimes (line) for QDs families.
			Triangles mark values obtained theoretically \cite{Gong2011} (\textcolor{gong}{$\blacktriangle$}) and experimentally \cite{Tomimoto2007} (\textcolor{tomimoto}{$\bm{\triangledown}$}) for other InAs/InP QDs.}
	\end{figure}
	The emission from all maxima has been investigated in the time domain by means of TRPL.
	Then, recorded PL decays were analyzed employing the PTI FelixGX software by Photon Technology International that utilizes the maximal entropy method (MEM) \cite{JaynesPR1957, JaynesPR1957a}, which among many applications has proved effective in multi-exponential luminescence decay reconstruction \cite{Livesey1987,Kumar2001}.
	It allows one to extract photoluminescence lifetimes without initial knowledge on the number of underlying processes, and thus decay components.
	Additionally, it does not introduce artificial features into the distribution of lifetimes, which are not supported by the data, if a reliable model of experimental noise is used.
	The analysis yielded the most probable distributions of PL decay constants $\tau_i$, which through the formula
	\begin{equation}
		I(t) = I(0) + \sum_{i=1}^N A_i \exp\lr*{ -\frac{t}{\tau_i}}
	\end{equation}
	reconstruct the experimental curves with the level of certainty given by the amplitude of experimental noise.
	Here, $N$ is the number of components in the quasi-continuous distribution of lifetimes describing the decay, and $A_i$ are the corresponding amplitudes.
	The results of this analysis are presented in \subfigref{Decays_and_MEM}{a}, where the TRPL traces collected for A--F luminescence peaks in the low excitation-power regime are shown along with the reconstructed curves.
	Via MEM reconstruction one obtains a distribution of times of nonzero width, which, if not really present in the system, represents the incomplete information carried by a noisy data.
	However, given a set of well-resolved peaks of amplitudes $A_i$, it is valid to treat their means as single representative values.
	Lifetimes presented with open squares and circles in \subfigref{Decays_and_MEM}{b} result from such an analysis.
	For the highest-energy peaks, we deal with two-exponential decays, where the shorter lifetime corresponds to recombination of excitons confined in a well-formed QD, while the longer one most probably comes from electron-hole pairs weakly bound on small WL width fluctuations.
	Sizes of symbols mark the amplitudes of each of decay components, and vanishing of the longer one with the number of MLs is present, which is reasonable, as no multiple-ML-wide fluctuations are expected. 
	
	Investigated QDs exhibit a dispersion of the $t_1$ PL lifetimes, which is different from what is expected and typically observed for more commonly studied InAs/GaAs QDs.
	In the latter, confinement characteristics for both types of carrier are comparable, with a weak tendency of hole wave function to leak into the barrier.
	Additionally, such QDs are typically of in-plane sizes not exceeding \SI{\sim30}{\nano\meter}, which results in strong exciton confinement \cite{EfrosFTP1982}, characterized by a weak impact of electron-hole Coulomb interaction, energy of which is much smaller than single-particle level spacing.
	In this regime the oscillator strength is simply defined by the overlap of electron and hole envelopes, which is close to 1 for a range of typical QD geometries.
	Thus, the observed dispersion of radiative lifetimes is caused mostly by the direct dependence of the latter on transition energy, which may be found in the dipole approximation as $\propto E^{-2}$.
	
	Here we deal with radiative lifetimes of nearly no dispersion, which places investigated QDs in the middle between above-mentioned InAs/GaAs dots, and those grown in InP reported previously \cite{Gong2011}.
	In the latter, the main source of radiative-lifetime dispersion for excitons has been found to come from the single-particle confinement characteristics, which are inversed in InP as compared to the GaAs matrix.
	In InP matrix, the electron is the particle that is more prone to leak into the barrier, as it experiences weaker confinement than holes, which is highly pronounced for flat QDs.
	The impact of electron leakage on exciton oscillator strength and lifetime in the InAs/InP QD system has been studied theoretically \cite{Gong2011}, and for completeness we plot the corresponding values with full triangles in \subfigref{Decays_and_MEM}{b}.
	It was found that oscillator strength increases as electron wave function is increasingly better confined within the QD, i.\,e., for higher dots.
	For example, for a lens-shaped QD with diameter of \SI{25}{\nano\meter} and height $H=\SI{2.5}{\nano\metre}$ only \SI{53.5}{\percent} probability of finding the electron inside the QD was reported, while for $H=\SI{5.5}{\nano\metre}$ this value increases to \SI{81.2}{\percent} \cite{Gong2008}.
	Such decrease of lifetimes with decreasing energy was also observed experimentally \cite{Tomimoto2007} for InAs/InP QDs and is presented with open triangles in \subfigref{Decays_and_MEM}{b}.
	
	Due to very small height of QDs investigated here, especially those from families A and B, we also deal with the reduction of oscillator strength caused by the weakly localized electron wave function.
	Contrarily, the hole is strongly localized within the dot, which reduces the overlap of their envelopes.
	Within the strong confinement limit this would lead to a direct reduction of coupling to light.
	With increasing QD height, the electron wave function tends to be more localized in the dot and its overlap with the hole becomes more complete, which reduces the observed PL lifetime.
	While this factor is common for all InAs/InP QDs, in the case of those studied here, it does not lead to expected inverted PL lifetime trend.
	Instead of this, a plateau is found, which means that another mechanism has to play a significant role.
	
	We find it by investigating the details of exciton eigenstates comprised of Coulomb-correlated electron-hole configurations.
	Typically, any admixtures of higher bright single-particle configurations into the exciton ground state result in the increase of oscillator strength.
	This purely quantum effect can be intuitively understood as the ability of the exciton to recombine simultaneously via each of the superposed electron-hole pair states.
	For large QDs, in which level spacing for electrons and/or holes is low enough to be comparable with $\SI{\sim 20}{\milli\electronvolt}$ energy of their Coulomb interaction, this is the main source of lifetime reduction.
	While the dots considered here are flat, they are also relatively large in-plane.
	Additionally, which may seem counter-intuitive, we find that reduced height results in stronger electron-hole Coulomb correlations and as a result in larger admixtures of higher-energy configurations to the exciton ground state.
	For a fixed in-plane size of $L=\SI{45}{\nano\metre}$ the admixture of higher electron-hole configurations if reduced from \SI{16}{\percent} to \SI{5.5}{\percent} when the height changes from 1 to 7~MLs.
	This may be understood, as the stronger the confinement along the growth axis, the more carrier wave functions are forced to penetrate the whole available in-plane area of the QD, which effectively increases the volume available for the exciton.
	This effect, pushing the dispersion of lifetimes in discussed QDs back towards the one known from InAs/GaAs dots underlies the observed approximately flat plot of lifetimes versus emission energy.
	
	\subsection{Emission from single QDs}
	\begin{figure}[!tb]
		\begin{center}
			\includegraphics[width=\columnwidth]{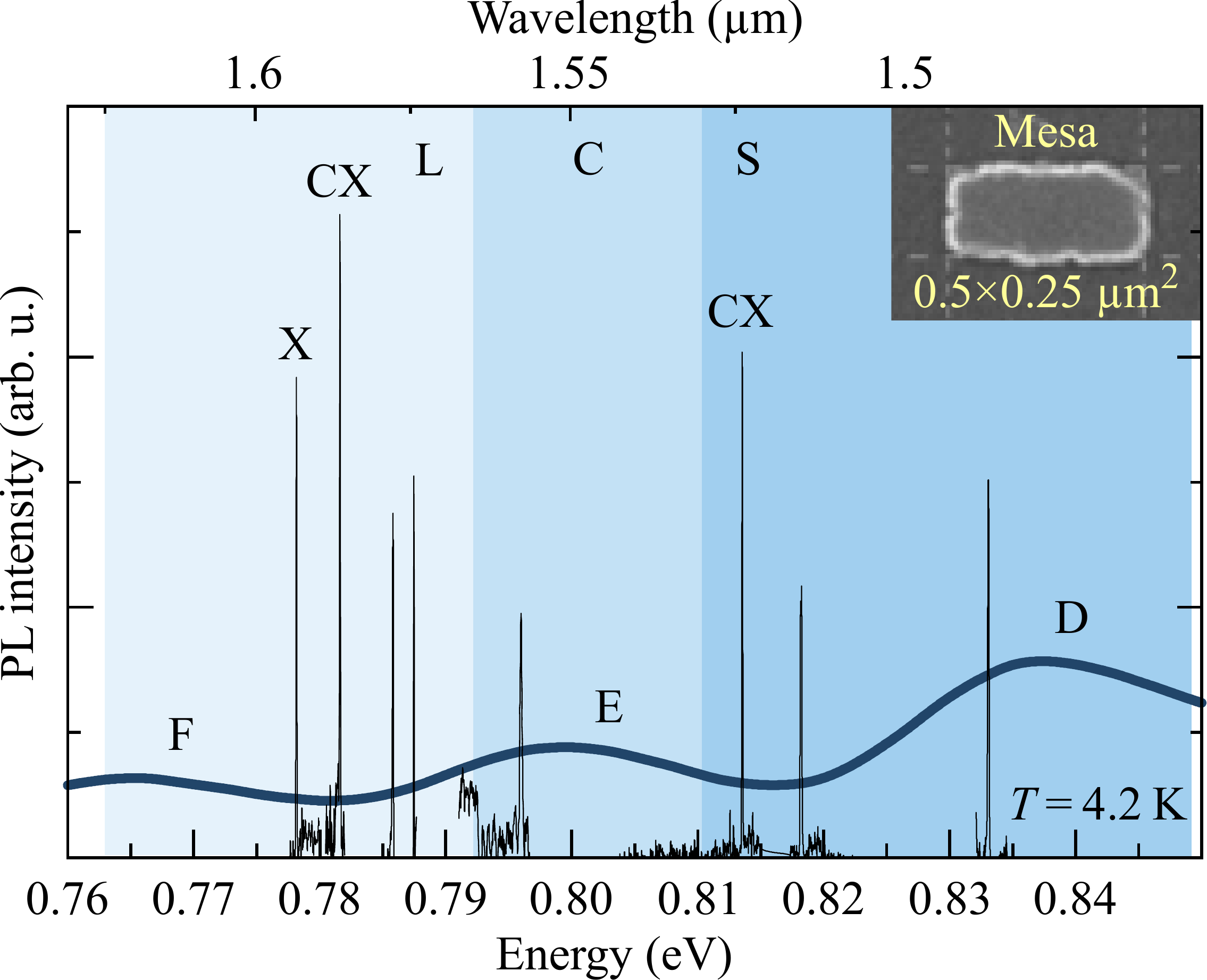}
		\end{center}
		\vspace{-0.8em}\caption{\label{fig:uPL_examples}%
			Part of the \si{\micro}PL spectrum from a mesa with QDs showing single emission lines observed at the third telecom window.}
	\end{figure}
	To deepen the understanding of the system in question, and to check the applicability of QDs toward the single-photon emission from spatially well-resolved emitters, we have performed optical experiments on spatially and spectrally isolated QDs.
	For better spatial separation of single emitters, the sample was processed with electron-beam lithography and etching, which left the array of mesa structures containing QDs.
	This enabled identification of exciton complexes and determination of their properties.
	An example of a single mesa is shown in the inset to \figref{uPL_examples}.

	We begin with presenting in \figref{uPL_examples} the results of \si{\micro}PL experiments performed on various mesas, which show intensive, well background-isolated emission lines.
	We attribute them to recombination of neutral exciton (X) and biexciton (XX) complexes as well as to charged excitons (CX), each coming from a single QD.
	Spectral positions of lines are compared with the PL spectrum from the ensemble of dots, showing that they belong to the D, E, and F families of QDs emitting in the L, C, and S telecom bands, respectively.
	We note that lines within L band are at the edge of efficiency for InGaAs-based multichannel array detector utilized here ($\SIrange{\sim1.6}{1.65}{\micro\meter}$), so single lines may also be present beyond this cut-off wavelength.
	Further identification of the emission lines shows that most of the dots emit due to recombination of charged exciton complexes (trions), and only a few of them show characteristics of the X and XX recombination.
	In this section we present examples of \si{\micro}PL investigations for both neutral and charged complexes.

	\begin{figure}[!tb]
	\begin{center}
		\includegraphics[width=\columnwidth]{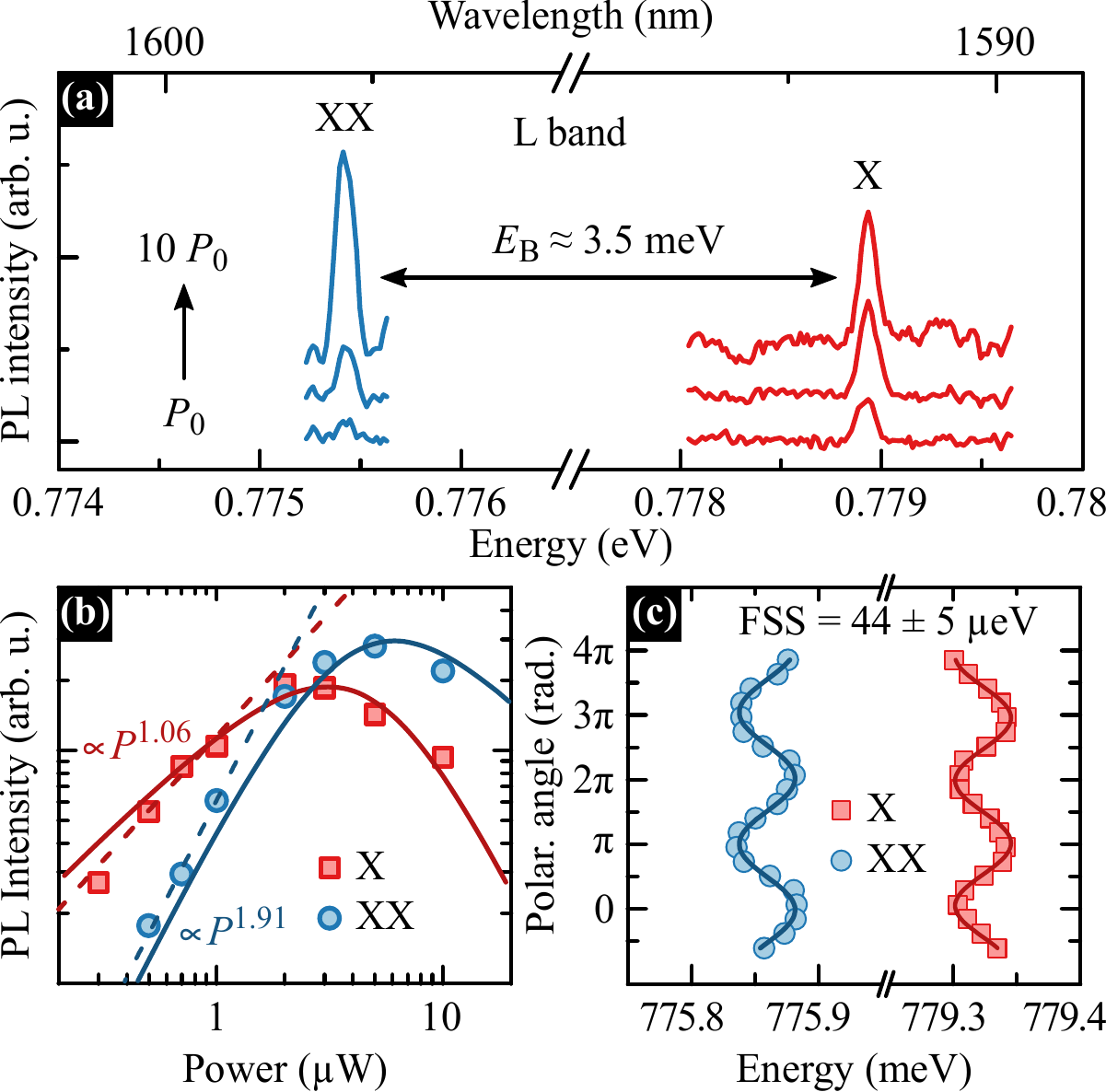}
	\end{center}
	\vspace{-0.8em}\caption{\label{fig:uPL_X_XX}%
		Exciton (X) and biexciton (XX) emission at L band.
		Measured biexciton binding energy is \SI{\sim3.5}{\milli\electronvolt}.
		$P_0=\SI{0.5}{\micro W}$
		(b) Power dependence of the \si{\micro}PL line intensities fitted with three-level rate-equation model shows a correlation between X and XX lines.
		(c) Polarization-resolved \si{\micro}PL reveals exciton fine structure splitting of $E_{\mr{FSS}} = 44 \pm 5$ \si{\micro\electronvolt}.}
	\end{figure}
	A pair of lines identified as the XX-X recombination cascade is analyzed in \figref{uPL_X_XX}.
	First, \si{\micro}PL spectra taken under various excitation powers are shown in \subfigref{uPL_X_XX}{a}.
	From spectral positions of lines we determine the biexciton binding energy of about $E_{\mr{XX}}-E_{\mr{X}} \simeq \SI{3.5}{\milli\electronvolt}$.
	The XX-X cascade was identified based on the power-dependent \si{\micro}PL, shown in \subfigref{uPL_X_XX}{b}, and polarization-resolved \si{\micro}PL presented in \subfigref{uPL_X_XX}{c}.
	Both the three-level rate-equation model \cite{Sek2010} [solid lines in \subfigref{uPL_X_XX}{b}] and anti-phase of energy oscillations in \subfigref{uPL_X_XX}{c} confirm correlation between observed lines.
	Additionally, in the low-power regime, the power dependence of X ($I_{\mr{X}}$) and XX ($I_{\mr{XX}}$) line intensities is well described with power laws: $I_{\mr{X}}\propto P^{1.06}$ and $I_{\mr{XX}}\propto P^{1.91}$, where $P$ is the excitation power.
	This is in a good agreement with expected approximately linear power dependence for X and quadratic for XX.
	In the polarization-angle domain, sinusoidal fits to the traces of emission energy yield the exciton fine structure splitting of $E_{\mr{FSS}} =  44 \pm 5$ \si{\micro\electronvolt}.

	\begin{figure}[!tb]
	\begin{center}
		\includegraphics[width=\columnwidth]{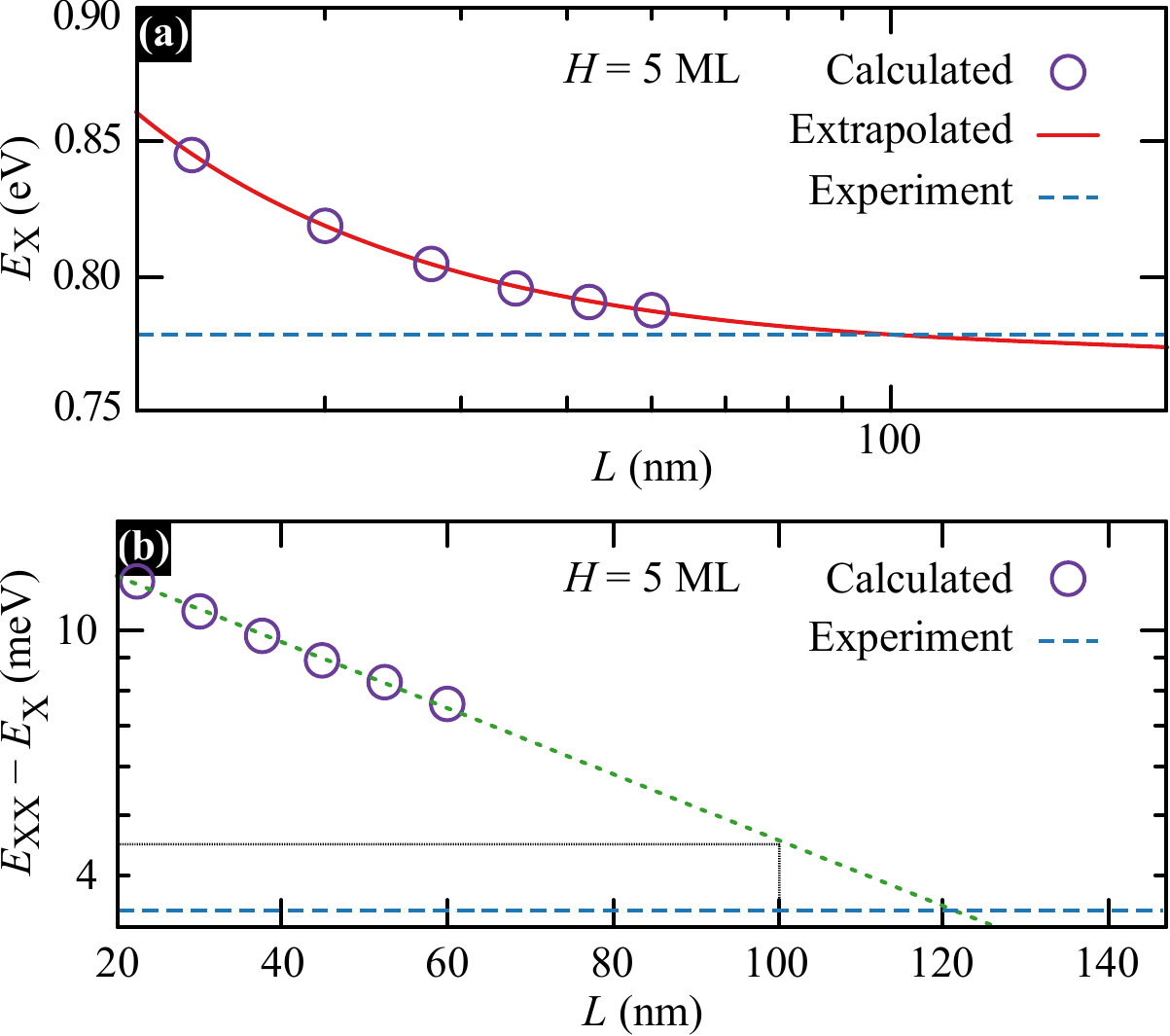}
	\end{center}
	\vspace{-0.8em}\caption{\label{fig:uPL_X_XX_2}%
		(a) Dependence of calculated exciton energy on QD in-plane size $L$ (circles) with extrapolation (solid line), and the value of X line energy from experiment (dashed line).
		(b) Dependence of calculated biexciton binding energy on QD length (circles) with approximate extrapolation (dashed line) and experimental value (dashed line).}
	\end{figure}
	The X-line energy places the emitting QD in the low-energy tail of the F family of QDs.
	Thus, properties different from those observed for the central part of the peak with the most typical QDs are expected.
	In particular, for a QD from this family to emit at such a long wavelength, the in-plane size has to be much higher than the average.
	This makes the direct modeling of such a QD numerically unfeasible.
	However, based on the very smooth dependence of basic electronic and optical properties of modeled QDs on their size, we can extrapolate results of our calculations to cover also such a nonstandard QD.
	To this end, we use the results of a series of calculations for QDs with in-plane size varying up to $L=\SI{60}{\nano\metre}$.
	First, we use the previously determined analytical approximation of exciton energy dependence on QD dimensions.
	Here, we use it to extrapolate the results beyond the range of in-plane sizes where the calculations were carried out.
	This is justified, as the formula used is not an arbitrary one, but the expected weakened $1/L^2$ dependence typical for QD confinement in general.
	The resulting trend is plotted in \subfigref{uPL_X_XX_2}{a} along with the measured energy of the X line (dashed line).
	As anticipated, the line is likely to come from a QD with large in-plane size of at least $L\sim\SI{100}{\nano\metre}$.
	Interestingly, it may be in fact arbitrarily large, as the position of observed line is close to the theoretically estimated asymptotic value for $L\to\infty$.
	
	Next, we calculate also the biexciton biding energy in the same range of QD sizes, and plot the results in \subfigref{uPL_X_XX_2}{b}.
	The points seem to follow a well-defined exponential curve, but as a nonvanishing binding energy in the quantum-well limit is expected, the dependence must have a nonzero asymptote.
	While the exact functional form is unknown, even the simplest estimation neglecting the expected curvature of the line in the given scale, marked with dashed line, yields for $L=\SI{100}{\nano\metre}$ a binding energy of $\SI{\sim4}{\milli\electronvolt}$, which agrees well with the experimentally found value.
	Thus, we conclude that the pair of lines is very likely to come from an atypically in-plane large QD.

	\begin{figure}[!tb]
		\begin{center}
			\includegraphics[width=\columnwidth]{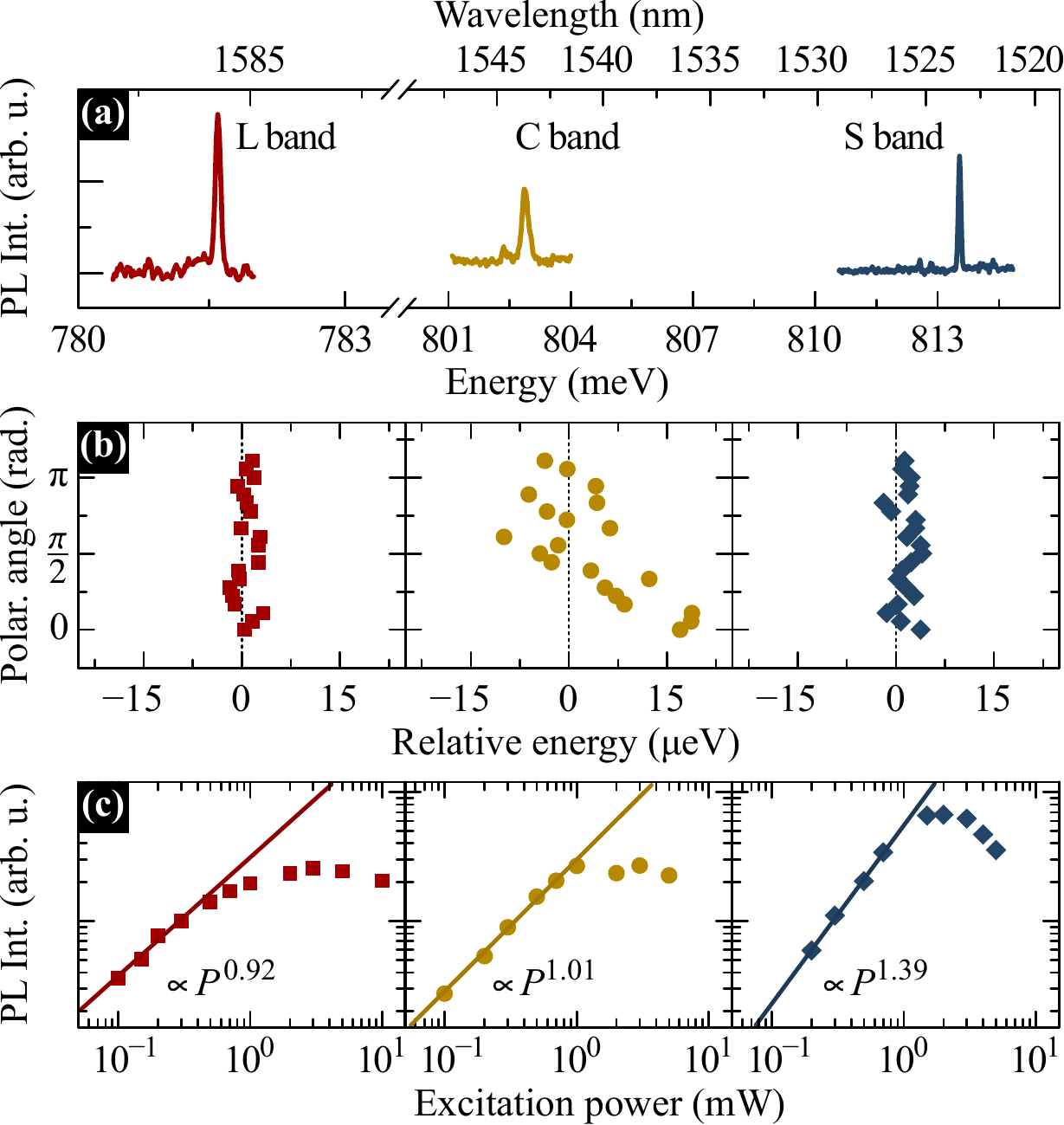}
		\end{center}
		\vspace{-0.8em}\caption{\label{fig:uPL_CX}%
			Emission from charged excitons (CX) in the third telecom window.
			(a)~\si{\micro}PL spectra.
			(b)~Polarization-resolved \si{\micro}PL showing no fine structure splitting.
			(c)~\si{\micro}PL power dependence (points) with fitted power-laws (lines).}
	\end{figure}
	Finally, in \figref{uPL_CX} we focus on charged complexes, optical signatures of which are found also within L, C and S bands.
	Trions, as spin singlet states, were identified based on the lack of a systematic variation in emission energy versus the angle of linear polarization of emission, as shown in \subfigref{uPL_CX}{b}, and close to linear power dependence of their intensity presented in \subfigref{uPL_CX}{c}.
	Additionally, observed CX lines exhibit no phase correlation to other nearby-lying lines within the biexciton binding energy range. This spectral isolation strengthens their assignment as charged complexes.

	\begin{figure}[!tb]
		\begin{center}
			\includegraphics[width=\columnwidth]{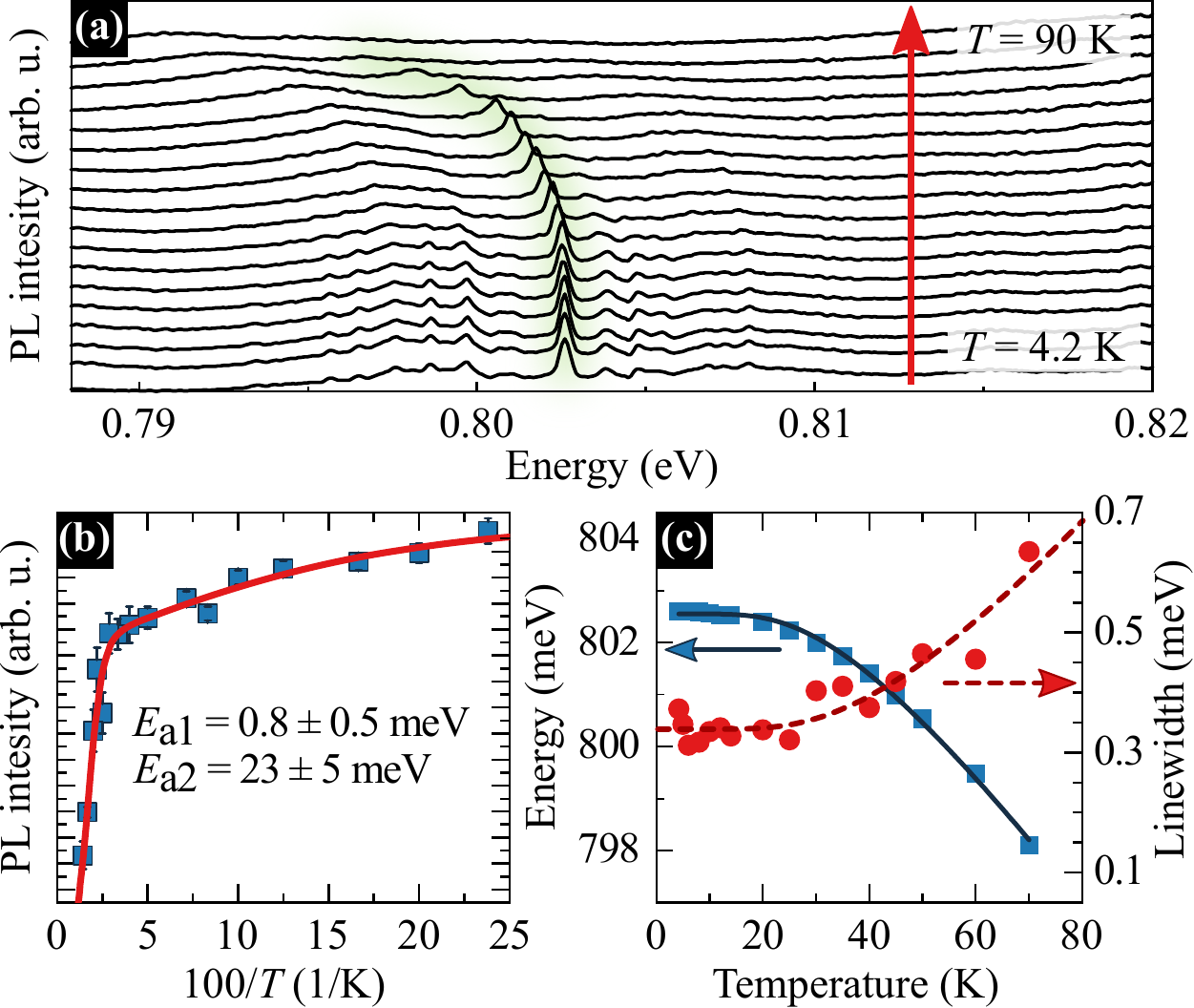}
		\end{center}
		\vspace{-0.8em}\caption{\label{fig:uPL_CX_Temp}%
			Temperature dependence of emission for charged exciton from a 5ML-high QD.
			(a) Stacked \si{\micro}PL spectra.
			(b) Temperature dependence of the \si{\micro}PL intensity (points) with a fit (line).
			(c) Energetic shift (squares) and linewidth broadening (circles) with fits (lines).}
	\end{figure}
	The CX line emitting in the C band has been chosen for further \si{\micro}PL temperature-dependent studies reported in \figref{uPL_CX_Temp}.
	The decreasing with temperature \si{\micro}PL signal was visible up to $T = \SI{70}{\kelvin}$.
	The Arrhenius plot in \subfigref{uPL_CX_Temp}{b} reveals two thermally activated processes.
	From fitting the data with Eq.~\eqref{ArrhEq} with two activation energies, we obtain values of $ E_{\mr{a1}} = 1.0 \pm 0.5$ meV and $ E_{\mr{a2}} = 23 \pm 5$ meV.
	To interpret this result, it is vital to underline the difference in possible sources of PL quenching for the ensemble and single emitters.
	In the former case, the main source is related to thermal escape of carriers from the localization centers, such as QDs.
	Other thermal effects, like excitation of complexes to their higher orbital levels do not impact the integrated intensity from a PL band, as those excited states also emit there.
	On the other hand, if a single line is studied, excitation of a carrier complex to its higher state results in the emission occurring at a different wavelength that is not accounted for during the analysis of the emission intensity.
	As a result, quenching of a single \si{\micro}PL line reveals the relative energy of the first considerably bright excited state.
	The line investigated in \figref{uPL_CX_Temp} lies in the high-energy tail of the ensemble, thus most probably comes from a QD with a relatively small in-plane size.
	Calculation for a QD with $H=5\,\mr{ML}$ and $L=\SI{28}{\nano\metre}$ yielded the lowest-energy bright excited CX states at $\Delta E = \SI{18.5}{\milli\electronvolt}$ and $\SI{18.7}{\milli\electronvolt}$ with radiative lifetimes of \SI{2.7}{\nano\second} and \SI{2.1}{\nano\second}, respectively.
	Thus, the energy $ E_{\mr{a2}}$ from experimental data may be attributed to the transfer of emission to a higher orbital state.
	
	The thermal redshift of CX emission is presented with squares in \subfigref{uPL_CX_Temp}{c} and fitted with formula accounting for phonon-modes occupation \cite{ODonnell1991}.
	Increase of the latter enhances exciton-phonon interaction, influencing the chemical bonds and their energy and as a result it is the main factor leading to the the band-gap renormalization \cite{Allen1976,Allen1983}.
	Extracted average phonon energy is $E_{\mr{ph,E}}=9.24\pm0.39$\:meV.
	We note that in contrast to \subfigref{TempSeries_DTU}{c} where we compared the thermal energy shifts of ensemble emission with Varshni curves, on the level of single emitters we can observe footprints of exciton-phonon interaction, therefore we use here more appropriate formula to explain the observed behavior.
	
	The linewidth broadening $\varGamma(T)$ is presented in \subfigref{uPL_CX_Temp}{c}. The rather broad initial value of $\varGamma\left(\SI{4.2}{\kelvin}\right)=\SI{340}{\micro\electronvolt}$ points to the crucial impact of defects or deep charge traps in the vicinity of a QD affecting significantly the emission properties. The broadening raises up to \SI{630}{\micro\electronvolt} at $T=\SI{70}{\kelvin}$. The temperature dependence of linewidth was fitted with the formula that includes the contribution of thermally-activated phonon sidebands to the zero-phonon line \cite{Gammon1996,Moody2011},
	\begin{equation}
		\varGamma(T)
		= \varGamma \left(\SI{4.2}{\kelvin}\right) + a\,\left[ \exp\left( \frac{E_{\mr{ph,\varGamma}}} {k_BT} \right) - 1 \right],
	\end{equation}
	where $k_B$ is the Boltzmann constant, parameter $a=1.5 \pm 1.1$~meV and $E_{\mr{ph,\varGamma}}=11.5\pm3.4$~meV is an average energy of phonons.
	We note that average energies of phonons obtained both from energy shift $E_{\mr{ph,E}}=9.24\pm0.39$~meV and linewidth broadening $E_{\mr{ph,\varGamma}}=11.5\pm3.4$~meV agree well with one another and with previously reported value for InAs/InP stacked QDs \cite{Ishi-Hayase2007}.
	
	\section{Summary}\label{sec:summary}
	In conclusion, we have obtained InAs/InP QDs with low surface density grown by MOVPE and performed detailed and comprehensive studies of their morphological and optical properties, both experimental and theoretical.
	QDs divide into families of different heights, from one to a few MLs of material above the WL, which is reflected in their emission with well-resolved multiple peaks covering a broad spectral range including a third window of silica-based optical fibers.
	In time-resolved spectroscopy, we observe dispersion of radiative lifetimes different from previously observed in InAs/InP QDs.
	While these typically show lifetimes increasing with emission energy in contrast to the opposite trend found in InAs/GaAs QDs, the dots under investigation show nearly no dispersion, which places them in between in respect to an electron-hole confinement parameters. 
	With theoretical modeling, comprising multiband $\kp$ calculations of single-particle states and configuration-interaction method for neutral and charged excitons, we explain the observed dispersionless lifetimes as originating from the interplay of weak electron confinement (typical for InAs/InP) and strongly height-dependent exciton confinement regime.
	The latter is shown in the amount of higher single-particle eigenstates admixed to the exciton ground state, changing from \SI{5.5}{\percent} for 7\:ML to \SI{16}{\percent} for 1\:ML.
	Thus, the flatter the dot the weaker the exciton confinement regime, which may be understood in geometric terms as a result of higher effective in-plane size of a QD that is penetrated by carrier wave functions.
	At higher temperatures a redistribution of carriers among families of QDs takes place, in which we observe escape of both single particles and whole excitons resulting in PL quenching, as well as recapturing of holes by higher QDs leading to emission enhancement at moderate temperatures.
	Thanks to low surface density of QDs, we have observed XX-X cascade and charged-exciton emission from single emitters persisting up to $T=\SI{70}{\kelvin}$.
	With theoretical modeling, we have reproduced experimental findings like emission energies, radiative lifetimes, thermal activation energies for PL quench and enhancement, and biexciton binding energy, which allowed for a comprehensive and detailed study of the system.
	
	\begin{acknowledgements}
		We acknowledge support from the Polish National Science Centre under Grant No 2014/14/M/ST3/00821, Villum Fonden via the NATEC Centre (8692), and YIP QUEENs project (VKR023442).
		P.\,H. acknowledges financial support from the Polish budgetary funds for science in 2018-2020 via the ``Diamond Grant'' program, Grant No DI~2017~011747.
		Numerical calculations have been carried out using resources provided by Wroclaw Centre for Networking and Supercomputing (\url{http://wcss.pl}), Grant No.~203. We would like to thank Christian Schneider for mesa fabrication as part of the collaboration between Wroc\l{}aw University of Science and Technology and the University of W\"{u}rzburg within the International Academic Partnership project funded by the Polish National Agency for Academic Exchange.
		We thank Krzysztof Gawarecki for sharing his implementation of the $\kp$ method, as well as Janusz Andrzejewski and Grzegorz S\k{e}k for helpful discussions.
		
		P.\,H. and M.\,G. contributed equally to this work.
	\end{acknowledgements}
	
	\FloatBarrier
	%\bibliography{bibliografia} % arxiv OFF
	%\input{sftunnel.bbl} % arxiv ON

\end{document}